\documentclass[aps,prl,showpacs,twocolumn]{revtex4}
\usepackage{graphicx}
\usepackage{amsmath}
\usepackage{amsfonts}
\usepackage{color}
\usepackage[normalem]{ulem}

\begin{document}

\title{Fractional Quantum Hall Effect in Topological Flat Bands
with Chern Number Two}
\author{Yi-Fei Wang$^{1}$, Hong Yao$^{2,3}$, Chang-De Gong$^{1,4}$, and D. N. Sheng$^{5}$
}
\affiliation{$^1$Center for Statistical and Theoretical Condensed
Matter Physics, and Department of Physics, Zhejiang Normal
University, Jinhua 321004, China
\\$^2$Institute for Advanced Study, Tsinghua University, Beijing 100084, China
\\$^3$Department of Physics,
Stanford University, Stanford, California 94305, USA
\\$^4$National Laboratory of Solid State Microstructures
and Department of Physics, Nanjing University, Nanjing 210093, China
\\$^5$Department of Physics and
Astronomy, California State University, Northridge, California
91330, USA}
\date{\today}

\begin{abstract}
Recent theoretical works have demonstrated various robust Abelian
and non-Abelian fractional topological phases in lattice models with
topological flat bands carrying Chern number $C=1$. Here we study
hard-core bosons and interacting fermions in a three-band
triangular-lattice model with the lowest topological flat band of
Chern number $C=2$. We find convincing numerical evidence of bosonic
fractional quantum Hall effect at the $\nu=1/3$ filling
characterized by three-fold quasi-degeneracy of ground states on a
torus, a fractional Chern number for each ground state, a robust
spectrum gap, and a gap in quasihole excitation spectrum. We also
observe numerical evidence of a robust fermionic fractional quantum
Hall effect for spinless fermions at the $\nu=1/5$ filling with
short-range interactions.

\end{abstract}

  \maketitle

{\it Introduction.---} Topological states of matter have been the
focus of intensive studies since the discovery of the integer
quantum Hall effect (QHE)~\cite{Klitzing} and the fractional QHE
(FQHE)~\cite{Tsui}. The latter, occurring at fractional filling of
Landau levels (LLs), provides the first example of fractionalization
in two dimensions. The precise quantization of Hall conductance was
found to be directly connected to a topological invariant Chern
number~\cite{Thouless,Niu} soon after its experimental discovery.
The FQHE is further characterized by quasi-particles with fractional
charge~\cite{Laughlin} and fractional
statistics~\cite{Haldane2,Halperin} as well as topological
ground-state degeneracy~\cite{WenQiu}, which are manifestations of
its topological order \cite{wen-book}. The idea of flux attachment
and composite-particle theory has provided a simple but profound
picture of the FQHE~\cite{ZhangKivelson,Jain}.

Recently, a series of numerical works have demonstrated convincing
evidence of the Abelian~\cite{Sheng1,YFWang,Regnault2} and
non-Abelian FQHEs~\cite{YFWang2,Bernevig,Bernevig2} in topological
flat band (TFB) models~\cite{Wen} without an external magnetic
field. These TFB models, belonging to the topological class of the
well-known Haldane model~\cite{Haldane}, have at least one
topologically nontrivial nearly flat band with a Chern number $C=1$,
which is separated from the other bands by large
gaps~\cite{Wen,Fiete,Venderbos,Mueller}. This intriguing
fractionalization effect in TFBs without LLs, defines a new class of
fractional topological phases (also known as fractional Chern
insulators), and has stimulated a lot of recent research
activities~\cite{XLQi,Sondhi,Goerbig,Murthy,
parton,Neupert,Xiao,Venderbos2,Ghaemi,Yang}.

In contrast to the continuum model in a magnetic field where the
Chern number of a LL is always one, higher Chern numbers are
possible for nearly flat bands in lattice models~\cite{FWang}. The
Abelian and non-Abelian FQHE states found in $C=1$
TFBs~\cite{Sheng1,YFWang,Regnault2,YFWang2,Bernevig,Bernevig2}
generally have analogy with ones in the continuum LL, while FQHE in
TFBs with high Chern numbers might do not have such simple analogy;
thus new exotic topological states of matter might occur in these
TFBs~\cite{XLQi2}. Nonetheless, exotic FQHE in TFBs with high Chern
numbers has not been studied in microscopic models. In this Letter,
we aim to fill in the gap by demonstrating that exotic FQHE for both
hard-core bosons and interacting fermions can indeed be realized in
TFBs with $C=2$.

We first introduce a three-band triangular lattice tight-binding
model whose lowest TFB has Chern number $C=2$. Through extensive
exact diagonalization (ED) studies of hard-core bosons in the TFB,
we find convincing numerical evidence of the bosonic FQHE at the
filling $\nu=1/3$ in the $C=2$ TFB. This $1/3$ bosonic FQHE is
characterized by three-fold quasi-degeneracy ($d=3$) of ground
states on a torus, a fractional quantized Chern number for each
ground state, a robust spectrum gap, and a gap in quasihole
excitation spectrum. For the $1/5$ filling of spinless fermions in
the $C=2$ TFB, clear FQHE features are also observed. We discuss the
plausibility of understanding the exotic $\nu=1/3$ bosonic and
$\nu=1/5$ fermionic FQHEs in the $C=2$ TFB in terms of effective
two-component (or bilayer) FQHEs.

\begin{figure}[tb]
  \vspace{0.0in}
  \hspace{0.0in}
 \includegraphics[scale=0.24]{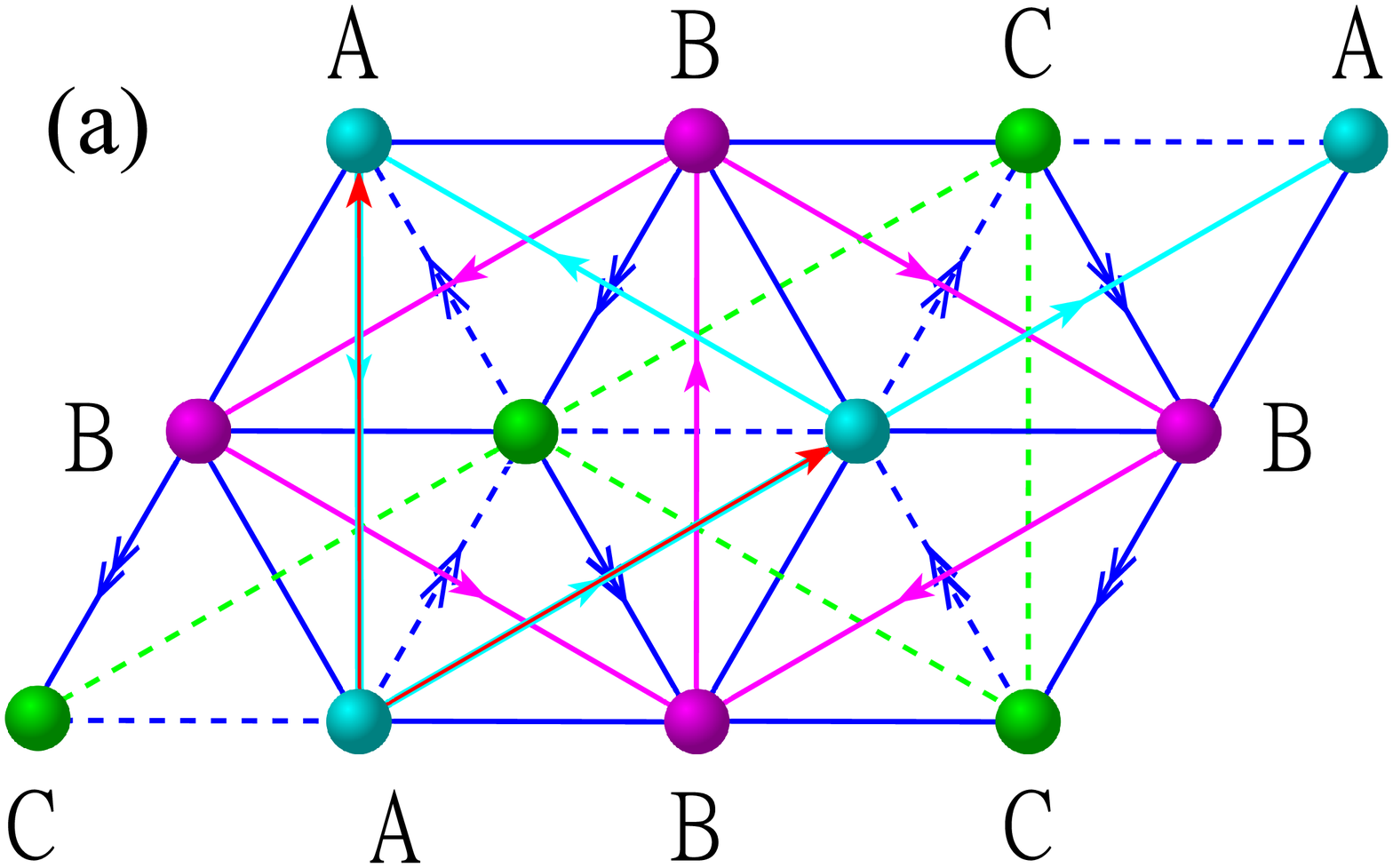}
  \hspace{-0.08in}
 \includegraphics[scale=0.40]{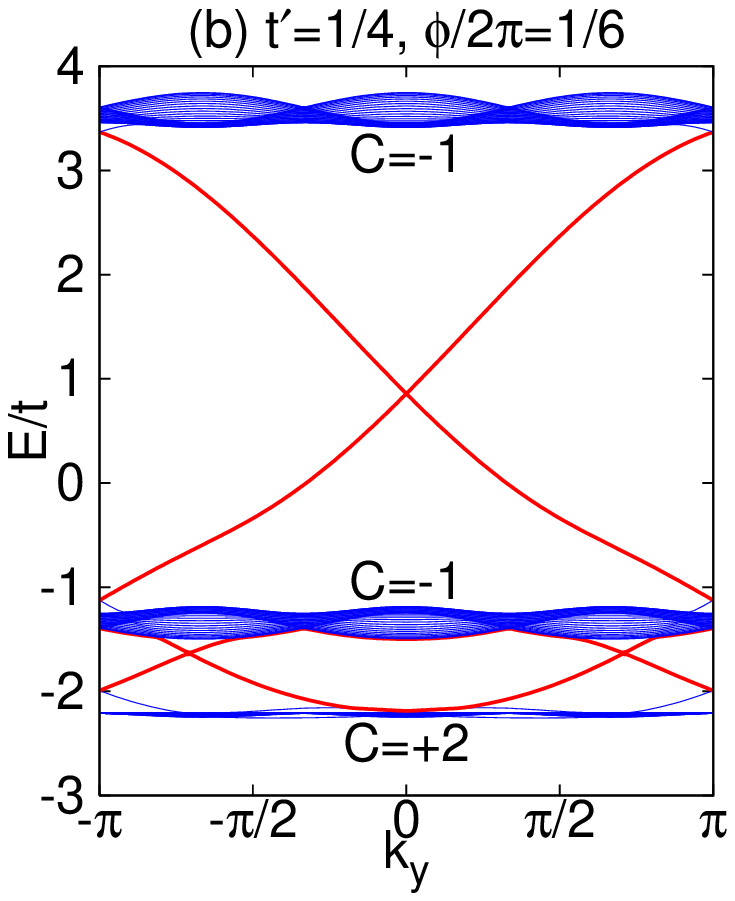}
  \vspace{-0.15in}
  \caption{(color online).
  (a) The three-band triangular-lattice model:
  The NN and NNN hopping amplitudes are positive (negative) along
  the solid (dashed) lines; The arrows represent the phases $\pm2\phi$
  (signs are represented by arrow directions)
  in the NN hoppings and $\pm\phi$ in the NNN hoppings.
  (b) Edge states of the triangular-lattice model in (a), and
  the lower TFB owns the Chern number $C=+2$.} \label{f.1}
\end{figure}

{\it Formulation.---}We introduce a three-band triangular-lattice
model of interacting hard-core bosons:
\begin{eqnarray}
H&=&\pm t\sum_{\langle\mathbf{r}\mathbf{r}^{ \prime}\rangle}
\left[b^{\dagger}_{\mathbf{r}^{ \prime}}b_{\mathbf{r}}\exp\left(i\phi_{\mathbf{r}^{ \prime}\mathbf{r}}\right)+\textrm{H.c.}\right]\nonumber\\
&&\pm t^{\prime}\sum_{\langle\langle\mathbf{r}\mathbf{r}^{
\prime}\rangle\rangle}
\left[b^{\dagger}_{\mathbf{r}^{\prime}}b_{\mathbf{r}}\exp\left(i\phi_{\mathbf{r}^{
\prime}\mathbf{r}}\right)+\textrm{H.c.}\right]\nonumber\\
&&+V_1\sum_{\langle\mathbf{r}\mathbf{r}^{ \prime}\rangle}
n_{\mathbf{r}}n_{\mathbf{r}^{\prime}}
+V_2\sum_{\langle\langle\mathbf{r}\mathbf{r}^{
\prime}\rangle\rangle}n_{\mathbf{r}}n_{\mathbf{r}^{\prime}}
\label{e.1}
\end{eqnarray}
where $b^{\dagger}_{\mathbf{r}}$ creates a hard-core boson at site
$\mathbf{r}$, $\langle\dots\rangle$ and
$\langle\langle\dots\rangle\rangle$ denote the nearest-neighbor (NN)
and the next-nearest-neighbor (NNN) pairs of sites, respectively
[Fig.~\ref{f.1}(a)], and $V_1$ and $V_2$ are the NN and the NNN
repulsions.

The triangular-lattice model has a unit cell of three sites, and
therefore has three single-particle bands. Here, we adopt the
parameters $t=1$, $t^{\prime}=1/4$ [the signs of hoppings are
descried in Fig.~\ref{f.1}(a)] and $\phi/2\pi=1/6$, such that a
lowest TFB of $C=2$ is formed with a flatness ratio (of the band gap
over bandwidth) of about $15$ [Fig.~\ref{f.1}(b)]. In our ED study,
we consider a finite system of $N_1\times N_2$ unit cells (total
number of sites $N_s=3\times N_1\times N_2$ and total number of
single-particle orbitals $N_{\rm orb}=N_1N_2$ in each band) with
basis vectors shown in Fig.~\ref{f.1}(a) and we use periodic
boundary conditions. We denote the boson numbers as $N_b$, and the
filling factor of the flat band is $\nu=N_b/N_{\rm orb}$. The
momentum vector $\mathbf{q}=(2\pi k_1/N_1,2\pi k_2/N_2)$ will be
denoted by a pair of integer quantum numbers  $(k_1,k_2)$. The
amplitude of the NN hopping $t$ is set as the unit of energy.

\begin{figure}[tb]
  \vspace{0.0in}
  \hspace{0.0in}
  \includegraphics[scale=0.58]{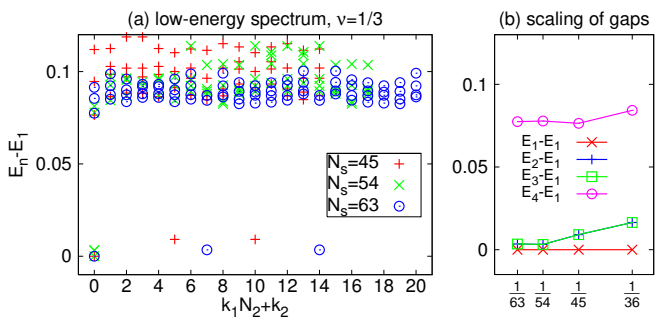}
  \vspace{-0.06in}
  \caption{(color online). The $1/3$ bosonic FQHE.
  (a) Low energy spectrum $E_n-E_1$ versus
  the momentum $k_1N_2+k_2$ of the $1/3$ bosonic FQHE phase for
  three lattice sizes $N_s=45$, $54$ and $63$ at $\nu=1/3$ filling with $V_1=V_2=0.0$.
  (b) Spectrum gaps versus $1/N_s$ for four lattice sizes.} \label{f.2}
\end{figure}

{\it The $1/3$ bosonic FQHE. (a) Low energy spectrum.---} We first
look at the low-energy spectrum for a finite lattice with $N_s=45$
($3\times3\times5$) sites at filling $\nu=1/3$ with $V_1=V_2=0.0$ as
shown by Fig.~\ref{f.2}(a). We denote $E_i$ as the energy of the
$i$-th lowest many-body eigenstate. The ground state manifold (GSM)
is defined as a set of lowest states with close energies well
separated from other excited states by a finite spectrum gap. For
the $\nu=1/3$ bosonic FQHE phase, two necessary conditions are
satisfied: a GSM with three quasi-degenerate ($d=3$) lowest
eigenstates ($E_3-E_1\sim0$); and the $d=3$ GSM being separated from
the higher eigenstates by a finite spectrum gap $E_4-E_3\gg
E_3-E_1$.

We have also obtained numerical results from other lattice sizes of
$N_s=36$ ($3\times3\times4$), $54$ ($3\times3\times6$) and $63$
($3\times3\times7$) around $V_1=V_2=0.0$. Similar to the FQHE in the
$C=1$ TFBs~\cite{Sheng1,YFWang,Regnault2}, if $(k_1,k_2)$ is the
momentum sector for one of the states in the GSM, we find that other
state in the GSM can be found in the sector $(k_1+N_b,k_2+N_b)$
[module $(N_1,N_2)$] demonstrating the momentum space translation
invariant as an emerging symmetry of the system. Indeed, for
$N_s=36$, $45$, $63$, the three GSs are in the (0,0), (1,0), and
(2,0) sectors, respectively; while for $N_s=54$, both $N_b/N_1$ and
$N_b/N_2$ are integers, and all three GSs are in the same (0,0)
sector [with very close energies as shown in Fig.~\ref{f.2}(a)].
Therefore, for each system size, there is an obvious GSM with
three-fold quasi-degenerate states, which is well separated from the
higher energy spectrum by a large spectrum gap. We have also
attempted a scaling plot of spectrum gaps for the four lattice sizes
[as shown in Fig.~\ref{f.2}(b)], which indicates that the spectrum
gap of the $1/3$ bosonic FQHE phase should survive in the
thermodynamic limit. We further find that this $\nu=1/3$ FQHE is
stable (with large spectrum gap and well-defined $d=3$ GSM) in the
presence of relatively weak repulsions ($V_1<0.5$ and $V_2<0.5$).

\begin{figure}[tb]
  \vspace{0.0in}
  \hspace{0.0in}
  \includegraphics[scale=0.58]{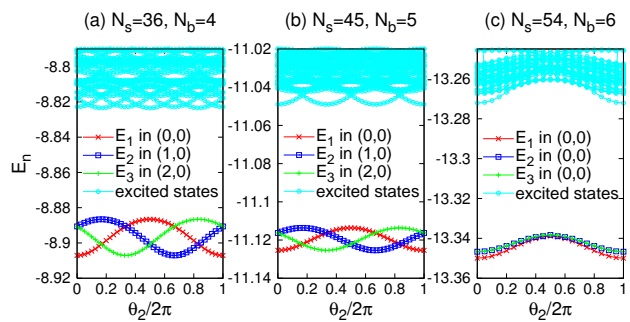}
  \vspace{-0.15in}
  \caption{(color online). The $1/3$ bosonic FQHE. Low energy spectra versus
  $\theta_2$ at a fixed $\theta_1=0$
  for three lattice sizes at $\nu=1/3$ filling with $V_1=V_2=0.0$:
  (a) $N_s=36$; (b) $N_s=45$; (c) $N_s=54$.} \label{f.3}
\end{figure}

\begin{figure}[tb]
  \vspace{-0.13in}
  \hspace{0.0in}
  \includegraphics[scale=0.62]{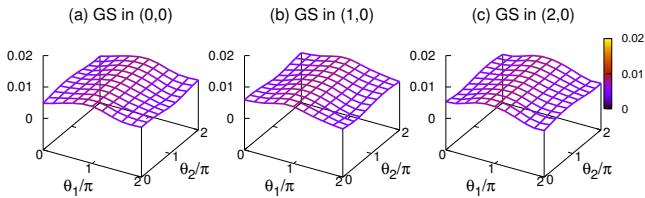}
  \vspace{-0.20in}
  \caption{(color online). The $1/3$ bosonic FQHE. Berry curvatures
  $F(\theta_1,\theta_2)\Delta\theta_1\Delta\theta_2/2\pi$
   at $10\times10$ mesh points for the GSM of the $N_s=45$ lattice at $\nu=1/3$ filling with $V_1=V_2=0.0$:
  (a) the 1st GS in (0,0) sector;
  (b) the 2nd GS in (1,0) sector;
  (c) the 3rd GS in (2,0) sector.} \label{f.4}
\end{figure}

{\it (b) Berry curvature and Chern number.---}The Chern
number~\cite{Thouless} (which is the Berry phase in units of $2\pi$)
of a many-body state is an integral invariant in the boundary phase
space~\cite{Niu,Sheng2}: $C={{1}\over{2\pi}}\int d\theta_1 d\theta_2
F(\theta_1,\theta_2)$, where two boundary phases $\theta_1$ and
$\theta_2$ are introduced as the generalized boundary conditions in
both directions, respectively. The Berry curvature is given by
$F(\theta_1,\theta_2)=\rm{Im} \left(\left\langle {{\partial
\Psi}\over{\partial\theta_2}}\Big{|}{{\partial
\Psi}\over{\partial\theta_1}}\right\rangle -\left\langle {{\partial
\Psi}\over{\partial\theta_1}}\Big{|}{{\partial
\Psi}\over{\partial\theta_2}}\right\rangle\right)$. For the GSM of
$1/3$ bosonic FQHE phase, the three GSs maintain their
quasi-degeneracy and are well separated from the other low-energy
excitation spectrum upon tuning the boundary phases, which indicates
the robustness of this FQHE phase (Fig.~\ref{f.3}). For each GSM of
$N_s=36,~45,~63$, the three states are found to evolve into each
other with level crossings when boundary phases are changed.
[Fig.~\ref{f.3}(a) and~\ref{f.3}(b)]. While for $N_s=54$, with all
three states of the GSM in the $(0,0)$ sector, through tuning the
boundary phases, each state evolves into itself without level
crossing [Fig.~\ref{f.3}(c)], consistent with the level repulsion
principle.

Furthermore, for the three GSs of the GSM in the (0,0), (1,0) and
(2,0) sectors of the $N_s=45$ case, the Berry curvatures in boundary
phase space (with $10\times 10$ mesh points) are shown in
Fig.~\ref{f.4}(a)-\ref{f.4}(c). The summation of Berry curvatures in
three sectors gives the integral Berry phase $4\pi$ (each GS
contributes an almost precisely quantized Berry phase $4\pi/3$ with
6-digit high accuracy), and thus the total Chern number of the $d=3$
GSM is $C_{\rm tot}=2$ which corresponds to a fractional quantized
Hall conductance of $2e^2/3h$ per GS.

\begin{figure}[tb]
  \vspace{0.0in}
  \hspace{0.0in}
  \includegraphics[scale=0.58]{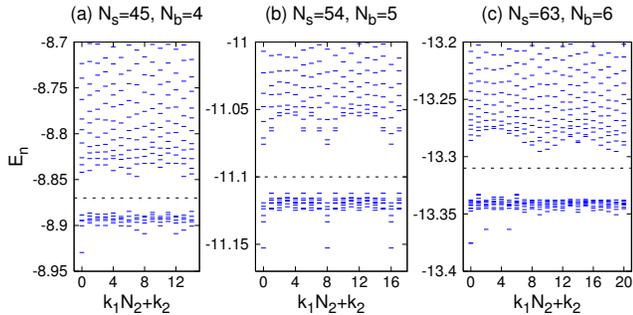}
  \vspace{-0.15in}
  \caption{(color online). The $1/3$ bosonic FQHE. Quasihole excitations for three lattice sizes with $V_1=V_2=0.0$:
  (a) $N_s=45$ and $N_b=4$;
  (b) $N_s=54$ and $N_b=5$;
  (c) $N_s=63$ and $N_b=6$;   } \label{f.5}
\end{figure}

{\it (c) Quasihole excitation spectrum.---}In order to investigate
the possible fractional statistics of the $1/3$ bosonic FQHE state,
we study the quasihole spectrum by removing one boson from the
$\nu=1/3$ filling. As shown in Fig.~\ref{f.5}(a), for the case of
$N_s=45$ and $N_b=4$, the quasihole spectrum exhibits a
distinguishable gap which separates $5$ lowest states in each
momentum sector from the other higher-energy states, and there are
$75$ low-energy quasihole states in total. This number of low-energy
quasihole states is consistent with the counting rule of splitting
one hole into three quasiholes (each with fractional charge $1/3$),
i.e. the quasihole-counting in Laughlin's $1/3$ fermionic FQHE
state, based upon the generalized Pauli
principle~\cite{Regnault2,Bernevig}. Similarly, for the $N_s=54$ and
$N_b=5$ case [Fig.~\ref{f.5}(b)], and the $N_s=63$ and $N_b=6$ case
[Fig.~\ref{f.5}(c)], there are also distinguishable spectrum gaps
and well-separated lower-energy quasihole manifolds.

\begin{figure}[tb]
  \vspace{0.0in}
  \hspace{0.0in}
  \includegraphics[scale=0.50]{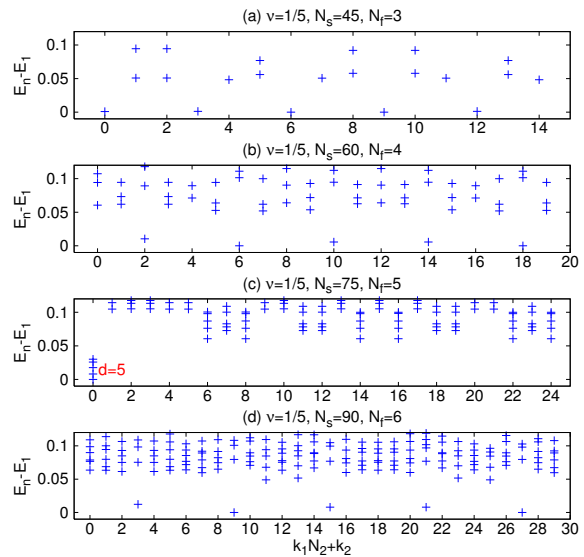}
  \vspace{0.0in}
  \caption{(color online).
  The $1/5$ ferminoic FQHE. Low energy spectrum $E_n-E_1$ versus
  the momentum $k_1N_2+k_2$ of the $1/5$ fermionic FQHE for
  four lattice sizes with $V_1=8.0$ and $V_2=V_3=1.0$:
  (a) $N_s=45$; (b) $N_s=60$;  (c) $N_s=75$; (d) $N_s=90$.
  The quasi-degeneracy has been labeled for the (0,0) sector of
  the $N_s=75$ case in (c).} \label{f.6}
\end{figure}

{\it The $1/5$ ferminoic FQHE.---}For the case of interacting
spinless fermions [where the bosonic operators in Eq.~(\ref{e.1})
are replaced by the fermionic ones] in the $C=2$ TFB, we have also
observed  FQHE features at the $\nu=1/5$ filling~\cite{note1} for
four different lattice sizes of $N_s=45$ ($3\times5\times3$), $60$
($3\times5\times4$), $75$ ($3\times5\times5$) and $90$
($3\times5\times6$). We denote the fermion numbers as $N_f$, and the
filling factor of the TFB is $\nu=N_f/N_{\rm orb}$. In contrast to
the $\nu=1/3$ bosonic FQHE around $V_1=V_2=0.0$, the onset of
fermionic FQHE features at $\nu=1/5$ needs finite values of
short-range repulsions ($V_1$, $V_2$ and a third-neighbor
interaction $V_3$), similar to the fermionic $1/5$ FQHE
~\cite{Sheng1} or the bosonic $1/4$ FQHE~\cite{YFWang} in the $C=1$
TFBs. At $\nu=1/5$, the topological ground-state degeneracy is
$d=5$: for $N_s=45$, $60$, $90$, the five GSs are in five different
sectors, e.g. the (0,0), (1,0), (2,0), (3,0) and (4,0) sectors for
$N_s=45$ [Fig.~\ref{f.6}(a), ~\ref{f.6}(b) and ~\ref{f.6}(d)]; while
for $N_s=75$, both $N_f/N_1$ and $N_f/N_2$ are integers, and all
five GSs are in the same (0,0) sector [with very close energies as
shown in Fig.~\ref{f.6}(c)]. Therefore, for each system size, there
is an obvious GSM with five-fold quasi-degenerate states, which is
well separated from the higher energy spectrum by a distinguishable
spectrum gap.

\begin{figure}[tb]
  \vspace{0.0in}
  \hspace{0.0in}
  \includegraphics[scale=0.58]{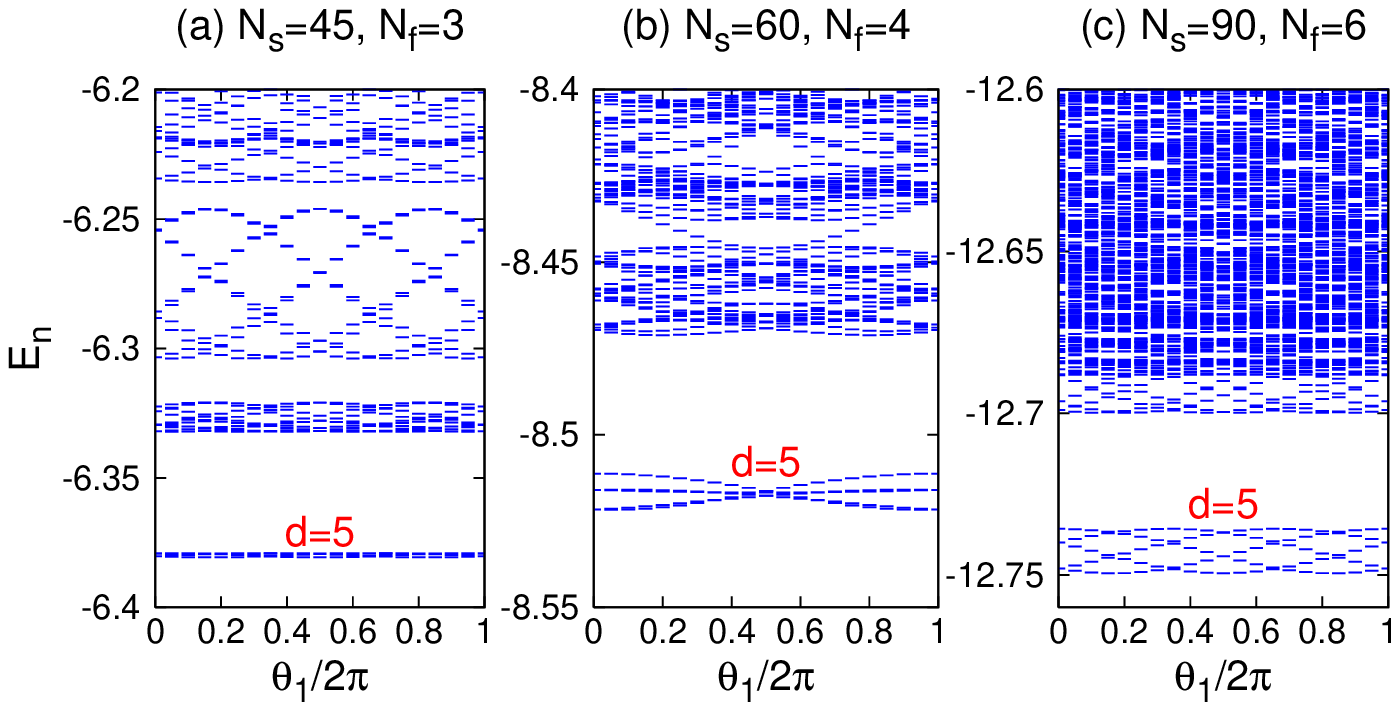}
  \vspace{-0.15in}
  \caption{(color online).
  The $1/5$ ferminoic FQHE. Low energy spectra versus
  $\theta_1$ at a fixed $\theta_2=0$
  for three lattice sizes at $\nu=1/5$ filling with $V_1=8.0$ and $V_2=V_3=1.0$:
  (a) $N_s=45$; (b) $N_s=60$; (c) $N_s=90$.
  The quasi-degeneracy of the GSM has been labeled.} \label{f.7}
\end{figure}

For the $1/5$ fermionic FQHE, the five GSs also maintain their
quasi-degeneracy and are well separated from the other low-energy
excitation spectrum when we tune the boundary phases
(Fig.~\ref{f.7}), indicating a possible robust topological phase.
Moreover, the $d=5$ GSM in the $1/5$ fermionic FQHE is found to
share a total Chern number $C_{\rm tot}=2$: e.g. for the $N_s=60$
case [Fig.~\ref{f.6}(b)], the summation of Berry curvatures in
$10\times10$ mesh points gives the Chern numbers $0.39629$,
$0.40975$, $0.38792$, $0.40975$ and $0.39629$ for the five GSs in
(0,2), (1,2), (2,2), (3,2) and (4,2) sectors, respectively; and thus
the total Chern number is found to be almost precisely $C_{\rm
tot}=2$ for this $d=5$ GSM, which implies that each GS supports a
fractional quantized Hall conductance of $2e^2/5h$.

{\it Concluding discussions.---} We find convincing numerical
evidences of $\nu=1/3$ bosonic FQHE in the $C=2$ TFB near
$V_1=V_2=0$. This odd-denominator bosonic FQHE phase is in stark
contrast to the $C=1$ TFB where the most robust bosonic FQHE occurs
at the $\nu=1/2$ filling for hard-core bosons with
$V_1=V_2=0$~\cite{YFWang}. It is desired to physically understand
the nature of this odd-denominator bosonic FQHE. Due to the absence
of higher topological ground-state degeneracy and the same quasihole
counting as Laughlin's $1/3$ fermionic FQHE, we believe that the
$1/3$ bosonic FQHE phase is of Abelian nature. Moreover, when we
treat the $C=2$ band as an effective two-component (or bilayer)
system~\cite{XLQi2} with the number of orbitals reduced to half for
each component (which doubles the effective total filling to 2/3),
the $1/3$ bosonic FQHE would be consistent, in terms of fractional
quasihole charge and ground state degeneracy, with Halperin's $mmn$
state (with $m=2$ and $n=1$) at the $2/3$ total filling. For the
$1/5$ filling of interacting spinless fermions in the $C=2$ TFB,
clear FQHE features have also been observed with a five-fold
degenerate GSM and a fractional quantized Hall conductance of
$2e^2/5h$ per GS, which could also be consistent with the Halperin
$332$ state. Nonetheless, the two ``components'' in a $C=2$ TFB are
mutually entangled, one may speculate that our states may be
different from the conventional bilayer $mmn$ states where the two
separated layers are only coupled by interaction. We believe that
more direct evidence is needed to verify the nature of the obtained
FQHE states, which calls for further studies in the future.

Other topological phases are also possible in other fractional
fillings. For the $\nu=1/4$ hard-core bosons, we have observed a
bosonic state with some topological features. However we can not
determine the nature of this $1/4$ bosonic state as its ground-state
degeneracy and the total Chern number varies with the particle
numbers which will be presented in the Supplementary
Material~\cite{supply}.

Although our results are obtained in a specific three-band
triangular-lattice model, one may speculate that such results,
especially the very robust $1/3$ bosonic state, might be universal
for generic $C=2$ TFB models~\cite{note2}. In the future, it would
be also highly interesting to demonstrate more exotic fractional
topological phases in microscopic TFB models with high Chern
numbers.

This work is supported in part by the NSFC of China Grant No.
10904130 (YFW), Tsinghua Startup Funds and the US NSF Grant No. DMR-0904264 (HY), the US DOE
Office of Basic Energy Sciences under Grant No. DE-FG02-06ER46305
(DNS), and the State Key Program for Basic Researches of China
Grants No. 2006CB921802 and No. 2009CB929504 (CDG).\\

{\it Note added.---}After the submission of the present Letter, a
few related works appeared very
recently~\cite{Bergholtz,SYang2,AMLauchli,Grushin,Sterdyniak}. Two
works have shown that flat bands with arbitrary high Chern numbers
can be systematically constructed using multi-layer lattice
models~\cite{Bergholtz,SYang2}. Another two works reported some
numerical evidence for a series of fermionic and bosonic fractional
incompressible states in a multi-layer kagome-lattice model with
high Chern numbers~\cite{AMLauchli,Sterdyniak}. These results
together with ours suggest some universality of FQHE in TFBs with
high Chern numbers.

\section*{Supplementary material for ``Fractional Quantum Hall Effect in
Topological Flat Bands with Chern Number Two''}

For the $1/4$ filling of hard-core bosons in $C=2$ TFBs, we have
also observed a bosonic state with some intriguing topological
features as described below.

\section*{The $1/4$ bosonic state}

\begin{figure}[!htb]
  \vspace{0.0in}
  \hspace{0.0in}
  \includegraphics[scale=0.50]{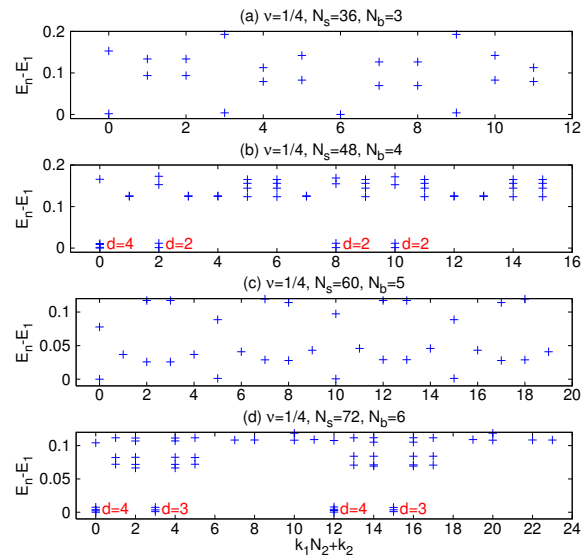}
  \vspace{0.0in}
  \caption{(color online).
  The $1/4$ bosonic state. Low energy spectrum $E_n-E_1$ versus
  the momentum $k_1N_2+k_2$ of the $1/4$ bosonic state for
  four lattice sizes with $V_1=8.0$ and $V_2=0.0$: (a) $N_s=36$; (b) $N_s=48$;  (c) $N_s=60$; (d) $N_s=72$.
  The quasi-degeneracy has been labeled for each sector
  with more than one GS.} \label{f.s1}
\end{figure}

We have also observed some FQHE features for hard-core bosons at the
$\nu=1/4$ filling from four lattice sizes of $N_s=36$
($3\times4\times3$), $48$ ($3\times4\times4$), $60$
($3\times4\times5$) and $72$ ($3\times4\times6$) at large $V_1$ and
zero or small $V_2$. In contrast to the bosonic $1/3$ FQHE around
$V_1=V_2=0.0$, the onset of FQHE features at $\nu=1/4$ needs a
finite value of $V_1$ ($>1.0$) similar to the fermionic $1/3$ FQHE
in the $C=1$ TFB. At $\nu=1/4$, the topological ground-state
degeneracy varies with the particle numbers: for the cases of odd
particle numbers, $N_b=3$ ($N_s=36$) and $N_b=5$ ($N_s=60$), the GSM
has $d=4$ quasi-degeneracy [Fig.~\ref{f.s1}(a) and ~\ref{f.s1}(c)];
for the case of $N_b=4$ and $N_s=48$, the GSM has $d=10$
quasi-degeneracy [Fig.~\ref{f.s1}(b)]; for the case of $N_b=6$ and
$N_s=72$, the GSM has $d=14$ quasi-degeneracy [Fig.~\ref{f.s1}(d)].
We can not predict what will happen at the thermodynamic limit based
on these size dependent results, but larger lattice sizes (e.g.
$N_b=8$ and $N_s=96$) are far beyond the capability of the current
ED method. However Fig.~\ref{f.s1} indicates that the spectrum gap
is decreased by a factor around $3$ ($2$) when going from $N_b=3$
($N_b=4$) to $N_b=5$ ($N_b=6$). This is in stark contrast to the
case of the $1/3$ bosonic FQHE state where the spectrum gap remains
roughly constant when the system size increases. The decrease of the
spectrum gap indicates a possible gapless state with a finite
compressibility in the thermodynamic limit.

\begin{figure}[!htb]
  \vspace{0.0in}
  \hspace{0.0in}
  \includegraphics[scale=0.58]{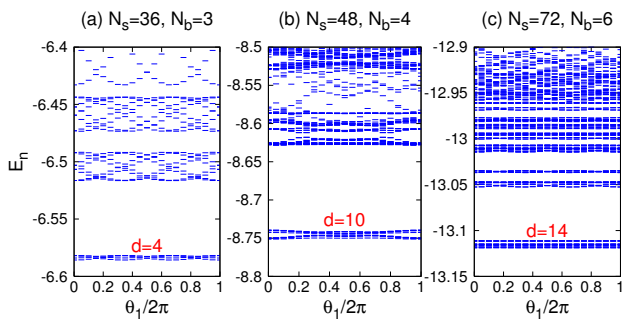}
  \vspace{-0.15in}
  \caption{(color online).
  The $1/4$ bosonic state. Low energy spectra versus
  $\theta_1$ at a fixed $\theta_2=0$
  for three lattice sizes at $\nu=1/4$ filling with $V_1=8.0$ and $V_2=0.0$:
  (a) $N_s=36$; (b) $N_s=48$; (c) $N_s=72$.
  The quasi-degeneracy of the GSM has been labeled.} \label{f.s2}
\end{figure}

For the $1/4$ bosonic state, the GSs also maintain their
quasi-degeneracy and are well separated from the other low-energy
excitation spectrum when we tune the boundary phases
(Fig.~\ref{f.s2}), indicating a possible robust topological phase.
Moreover, the $d=4$ GSM in the $1/4$ bosonic state at odd $N_b$'s is
found to share a total Chern number $C_{\rm tot}=2$: e.g. for the
four GSs of the $d=4$ GSM in the (0,0), (1,0), (2,0) and (3,0)
sectors of the $N_s=36$ case and the $N_s=60$ case, an integral
Berry phase $4\pi$ is obtained, thus the total Chern number of the
$d=4$ GSM is $C_{\rm tot}=2$ (each GS contributes a Berry phase
$\pi$ with 4-digit accuracy for $10\times10$ mesh points of the
$N_s=60$ case). For the GSM at $N_b=4$ and $N_s=48$, the $d=10$ GSM
is found to share a total Chern number $C_{\rm tot}=5$ (five GSs
with a Berry phase $4\pi$ each and the other five GSs with a Berry
phase $-2\pi$ each of 6-digit high accuracy for $10\times10$ mesh
points). For these cases, the quantization of Chern number
associated with each GS is $1/2$ in agreement with a possible FQHE
at $\nu=1/4$. However, for the GSM at $N_b=6$ and $N_s=72$, two
lowest GSs in the (0,0) sector [and the (2,0) sector] give the Berry
phases $2\pi$ and $-2\pi$, while the other ten GSs give a Berry
phase $\pi$ each, and thus the total Chern number is also $C_{\rm
tot}=5$. The variation of these quantized numbers may indicate a
finite compressibility of ground states in the thermodynamic limit.

After removing one boson from the $1/4$-filled $N_s=48$ case, i.e.
$N_b=3$, we have also found that the low-energy spectrum exhibits a
distinguishable gap which separates $6$ lowest states in each
momentum sector from the other higher-energy states. However for
other boson numbers, no distinguishable quasihole spectrum gap has
been found. The nature of such a state appears to be very
complicated, which we hope to address in the future using
alternative approaches.


\begin{thebibliography}{99}
\bibitem{note1} Very recently, a similar $1/5$ fermionic FQHE has
also been observed for two other $C=2$ TFB models in
Refs.~\cite{AMLauchli,Grushin}.
\bibitem{note2} Indeed, very recently, a similar $1/3$ bosonic FQHE has also been observed
for a bilayer kagome-lattice model in
Refs.~\cite{AMLauchli,Sterdyniak} and a two-orbital square-lattice
model (see an updated version of Ref.~\cite{SYang2}), supporting
this universality conjecture.
\end{thebibliography}
\end{document}